\documentclass[%
 aip,
 amsmath,amssymb,
 reprint,%
]{revtex4-1}

\usepackage{graphicx}
\usepackage{dcolumn}
\usepackage{bm}

\usepackage[utf8]{inputenc}
\usepackage[T1]{fontenc}
\usepackage{mathptmx}
\usepackage{etoolbox}
\usepackage{xcolor}
\usepackage{subfigure}
\usepackage{graphicx}
\usepackage[percent]{overpic} 
\makeatletter
\def\@email#1#2{%
 \endgroup
 \patchcmd{\titleblock@produce}
  {\frontmatter@RRAPformat}
  {\frontmatter@RRAPformat{\produce@RRAP{*#1\href{mailto:#2}{#2}}}\frontmatter@RRAPformat}
  {}{}
}%
\makeatother
\begin{document}

\preprint{AIP/123-QED}

\title{Barium Magnesium Alloy as Source of Atomic Ba for Ion Trapping}

\author{Jane Gunnell\textsuperscript{*}}
\email{janegunn@uw.edu}
\author{Thomas Griffiths}
\author{Boris B. Blinov}
\affiliation{Department of Physics, University of Washington, Seattle, WA 98122 USA}

\date{\today}

\begin{abstract}
Trapped atomic ion qubits exhibit long coherence times and high fidelity qubit state preparation, manipulation and detection, making them well-suited for scalable quantum computing applications. Among several atomic species used in quantum computing and other application, singly-charged ions of barium stand out due to their long wavelength transitions and the presence of very long-lived metastable internal states. However, elemental barium is a highly reactive metal making it experimentally difficult to work with when making atomic beam sources. In this paper, we demonstrate a method of using resistively heated ovens loaded with a barium magnesium alloy (BaMg) as a source of barium for ion traps. This alloy is not very chemically reactive and does not oxidize in air. We found that a sample of BaMg in a resistively heated oven produced barium vapor pressures on the same order as a metallic barium sample prepared the same way. Two separate ovens, one with a sample of BaMg and one with metallic barium, were used as source for an ion trap. We observed reliable trapping of $^{138}$Ba$^+$ ions both with the elemental barium source, and the BaMg source.
\end{abstract}

\maketitle

\section{Introduction}

 Quantum computers offer solutions to computational challenges that are difficult or impossible for classical computers\cite{quantumcomputinggood}. However, in order to achieve these advantages experimentally, physical systems must offer long coherence times, high-fidelity initialization, control and measurement, and scalable architectures\cite{textbook}. Trapped ions as qubits have been studied extensively and have emerged as one of the most effective candidates for quantum computing technology \cite{bible,progress}.

Trapped ions provide naturally identical qubits with exceptionally long coherence times and high fidelity state preparation and entangling gates\cite{coherence,fidelity,SPAM}. Among the ion species with high potential for quantum computing, barium ions offer several distinctive advantages that make them an especially attractive qubit platform. They possess a favorable electronic level structure with strong optical transitions in the visible and near-infrared, enabling efficient laser cooling, state detection, and coherent qubit control using readily available laser technology\cite{originalbarium,morebarium}. Barium ions also have a relatively large mass which reduces sensitivity to electric-field noise, leading to lower motional heating rates compared to lighter ion species and have long lived metastable states that are essential for quantum information processing \cite{atomicweight,bariumlifetimetheory,experimentalbalifetime,updatedlifetimeba}. Additionally, Barium ions lead in state preparation and measurement fidelities compared to other ion species \cite{highestSPAM}. 

Typically, ion traps use resistively heated ovens to produce neutral atoms, which are subsequently ionized \cite{atomicsourcesummary}. However, using this method with metallic barium can be experimentally difficult because of how quickly it oxidizes \cite{oxidize, borisbariumions}. Assembling the ion trap apparatus, including installing the atomic beam sources, is typically done in air, where barium would oxidize in a matter of minutes. An alternative technique used for creating neutral atoms or ions for trapping is laser ablation\cite{firstlaser, anotherlaser}. Here, to produce barium ions, a thin layer of BaCl$_2$ is typically used as a laser ablation target, which does not oxidize like elemental barium. However, when using laser ablation the position in which the laser hits the target needs to be adjusted frequently because to the thin layer of material and this can cause a variance in the atomic flux \cite{crystallaser}. Strides have been made to homogenize the atomic flux produced from these ablation targets \cite{ablationfix}, but resistively heated ovens still remain the most simple source of atomic flux for ion traps. 

In this paper, we propose and demonstrate a method for using a barium–magnesium (BaMg) alloy, which is stable in air, as a barium source in an oven for ion trapping. The BaMg sample used in this research was purchased from S.A. Materials and consists of 20\% barium and 80\% magnesium. A sample of the alloy is shown in Fig.1(a). It is a relatively light, white metal that is brittle but easily machined with hand tools, such as files and metal saws. We first tested the material to ensure that it produced barium vapor as it is heated in vacuum, as discussed in the following section. We then loaded BaMg and elemental barium into two separate ovens and put them in a vacuum chamber with an ion trap \cite{forkandring}. We established a trapping protocol to trap $^{138}$Ba$^+$ ions with the elemental barium oven and then used that same protocol for trapping using the BaMg oven. We found we were able to trap using the BaMg sample, and propose that it could be a viable substitute barium sources in resistively heated ovens or laser ablation. 

\begin{figure}[h]
    \centering
    \subfigure{
        \begin{overpic}[width=0.22\textwidth]{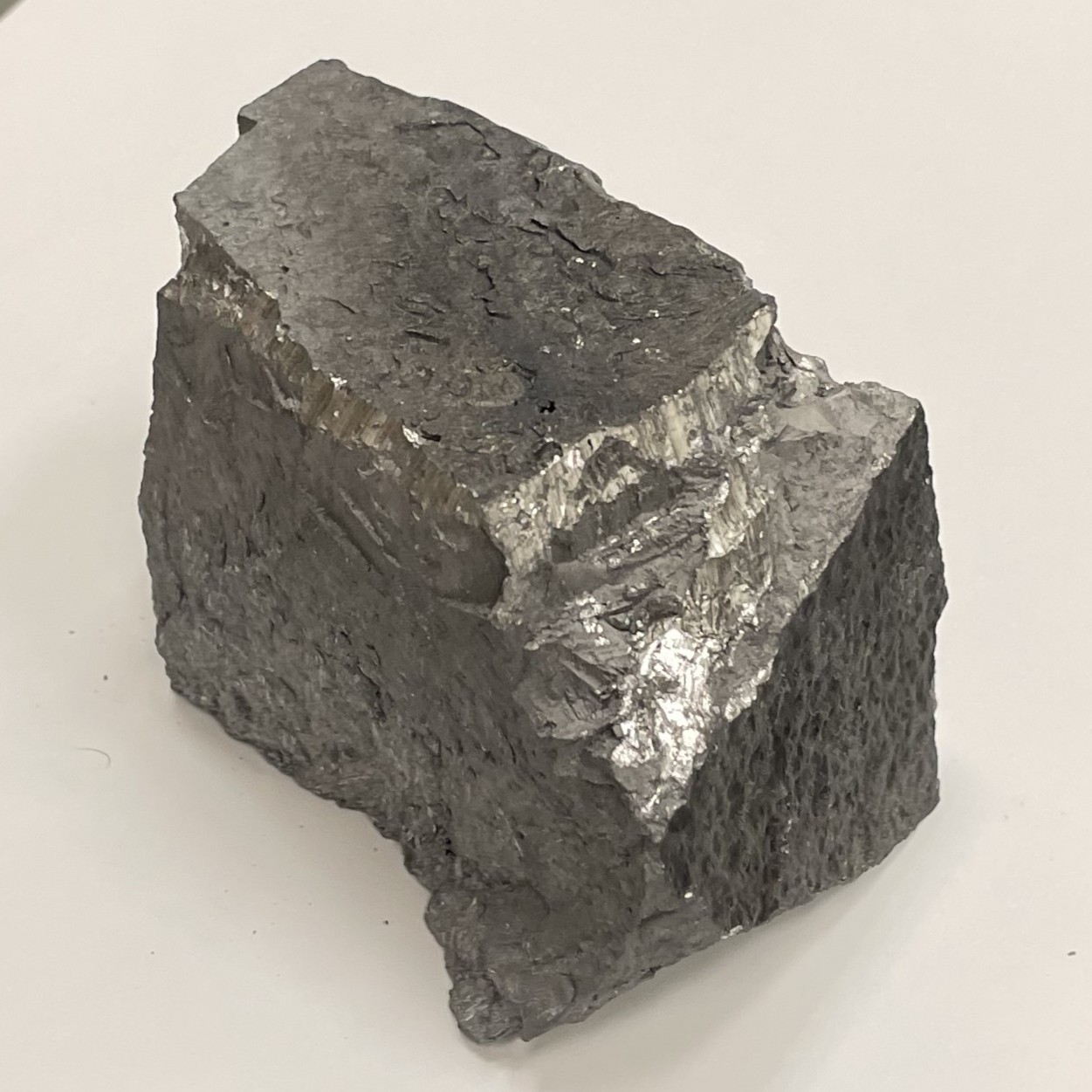}
            \put(2,90){\textbf{(a)}} 
        \end{overpic}
    }
    \subfigure{
        \begin{overpic}[width=0.22\textwidth]{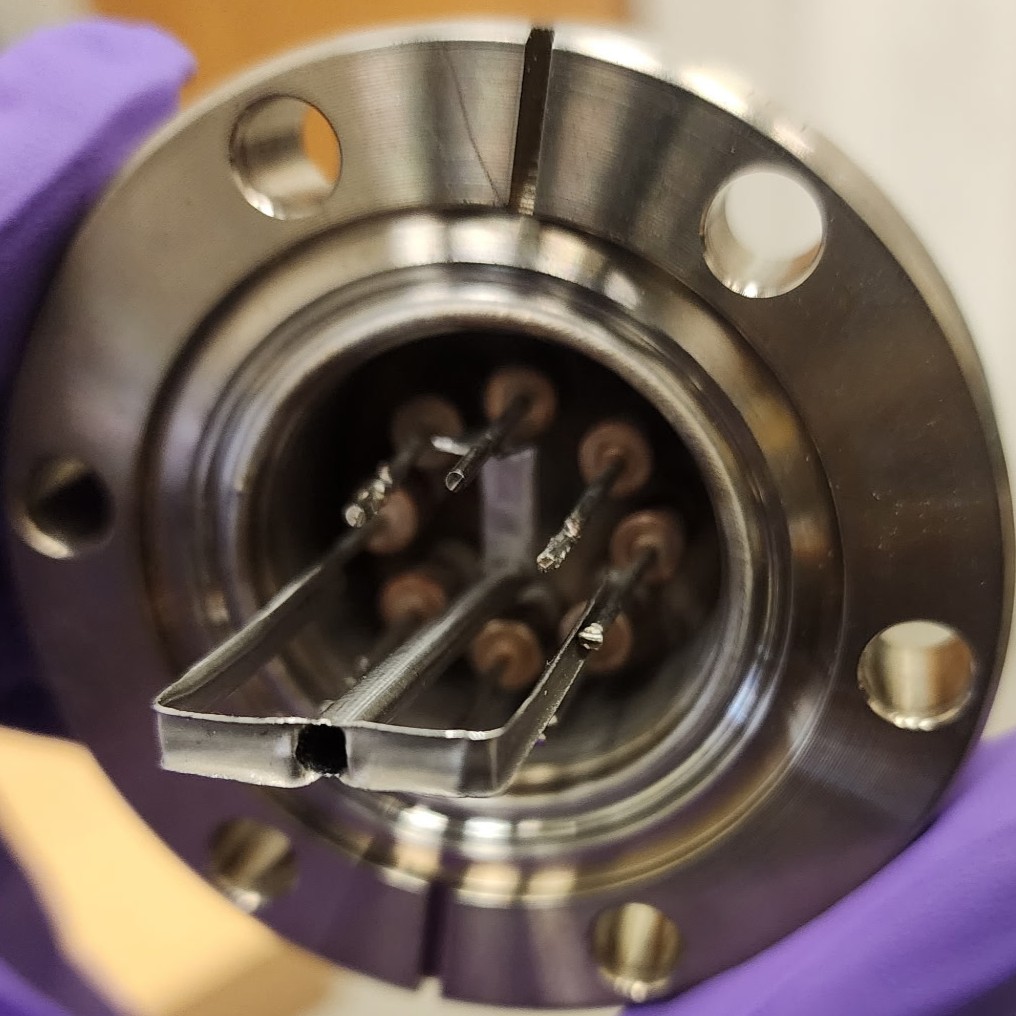}
            \put(2,90){\textbf{(b)}} 
        \end{overpic}
    }

    \caption{(a) A BaMg sample approximately 5~cm on each side. (b) An example of a stainless steel oven used in the preliminary testing of the BaMg as neutral Ba source.}
    \label{mgpic}
\end{figure}

\begin{figure*}[!ht]
\includegraphics[width=0.9\textwidth]{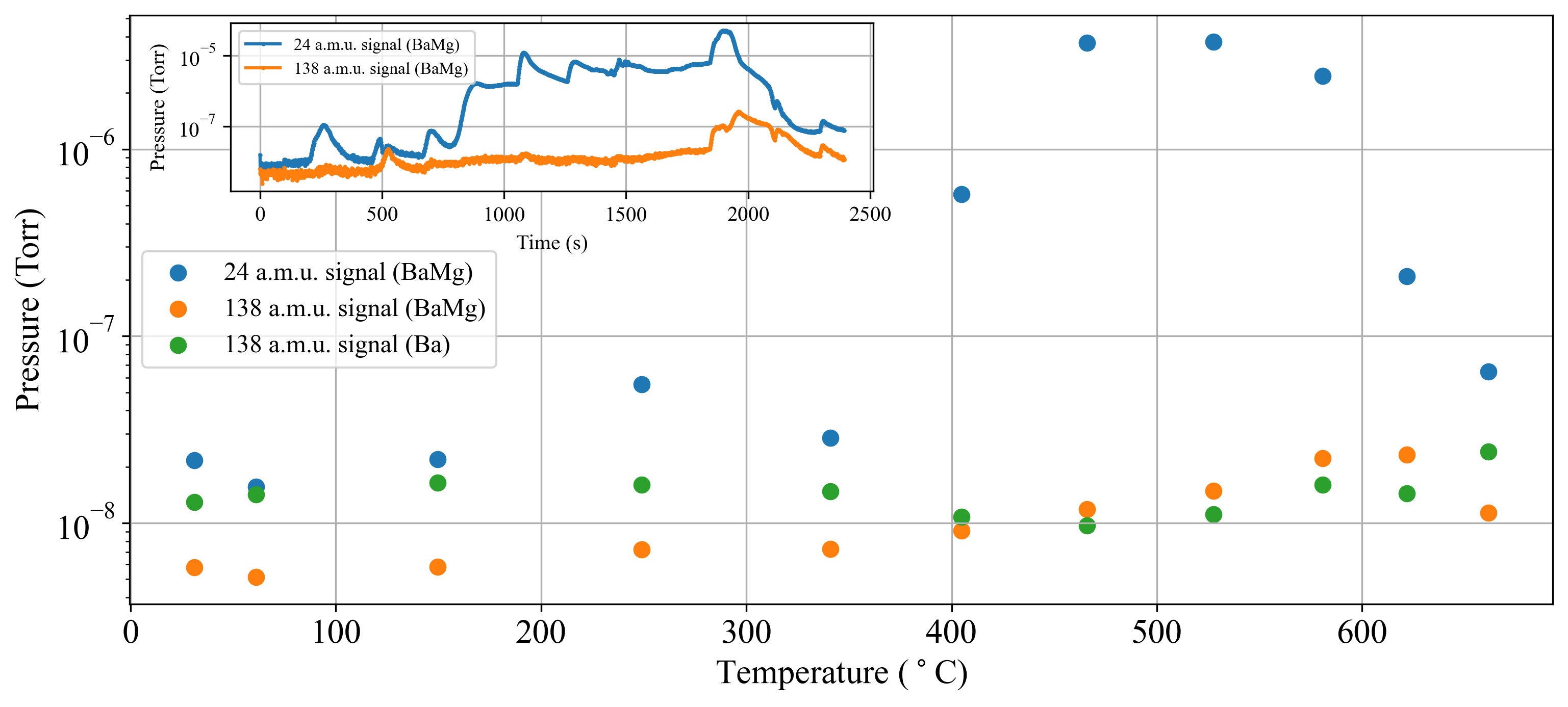}
\caption{Results from BaMg testing using a residual gas analyzer (RGA). The data points from the BaMg sample are averaged over three independent heating cycles. The 24 a.m.u. signal, corresponding to magnesium, exhibits a higher partial pressure at lower temperatures. In contrast, the 138 a.m.u. signal, corresponding to barium, increases with temperature but reaches a comparatively lower peak pressure. However, at higher temperatures, the 138 a.m.u. signal from the BaMg sample is on the same order of magnitude as that measured from a metallic barium sample. This suggests that BaMg can serve as a viable substitute for metallic barium as a barium source. The inset shows representative data recorded directly from the RGA during a single BaMg heating cycle. The oven current was increased in increments of 0.5~A to 1~A at several times during the heating cycle.}
\label{prelimtestresults}
\end{figure*}

\section{Oven flux testing}

For preliminary testing, a sample of BaMg was cut into small 1 to 2 mm pieces and loaded into a stainless steel oven, similar to one shown in Fig. 1(b), spot-welded to a ultrahigh vacuum (UHV) feedthrough. This oven was then installed to a vacuum system along with a Residual Gas Analyzer (RGA) placed such that the atomic flux from the oven would pass through the ionization volume of the RGA. A thermocouple was attached near the oven to monitor the temperature in the configuration described in ref. 15. The chamber was pumped to pressures below 10$^{-6}$ mbar and the oven was continuously heated by running a current through it. The current was increased in steps of 1~A while monitoring the RGA response at 24 a.m.u. and 138 a.m.u. corresponding to atomic Magnesium and Barium, respectively. In a separate experiment, an oven of the same design was loaded with a metallic barium sample \cite{BaSample}. The results of these measurements are summarized in Fig. \ref{prelimtestresults}.

The inset in Fig. \ref{prelimtestresults} shows an example of the data recorded from the RGA for one heating cycle of a BaMg sample. The dataset presented on the larger graph shows the pressure recorded at different oven temperatures averaged over three heating cycles for the BaMg sample, and a single heating cycle for the metallic barium sample. As expected, magnesium and barium are both detected by the RGA when heating the BaMg sample. The results show that the magnesium vapor increases to higher pressures at lower temperatures compared to the barium vapor. This is as expected considering that magnesium has a lower vaporization temperature \cite{vaporization}. When comparing the 138 a.m.u. to from the BaMg sample to metallic barium, at higher temperatures, the pressures were on the same order, which lead us to believe the BaMg sample could be used as a replacement barium source for an ion trap.

\section{Trapping Ba ions}

 Two ceramic ovens with tungsten resistive heaters, one with a sample of BaMg and one with a sample of metallic barium, were aligned in a vacuum chamber with a fork-and-ring trap~\cite{forkandring}. The geometry of this trap consists of a ring RF electrode and two grounded electrodes above and below the ring to confine the ion in the center of the ring. The ions were loaded into the trap by heating one of the atomic ovens and producing a stream of neutral atoms. The barium atoms were then ionized in a two-step photoionization process using a home built tunable 791~nm extended cavity diode laser (ECDL) and a pulsed 337.1~nm nitrogen laser \cite{photo}. The ions  were Doppler-cooled with homebuilt ECDL near 493~nm, with another ECDL (Toptica DL-Pro) near 650~nm used for repumping the ion out of a metastable dark state. Fluorescence from a trapped ion was monitored using an Andor Luca Electron-Multiplying Charge-Coupled Device (EMCCD) camera. A robust protocol for trapping with metallic barium was first established by repeatably trapping and dumping an ion. We found that we were able to trap a single $^{138}$Ba$^+$ ion on average every 124$\pm$17s, when running the oven at a current of 1.38 A. We then switched to working with the BaMg sample. We were first able to trap $^{138}$Ba$^+$ with the BaMg sample when running the oven at a higher current of 2~A. The current was then reduced to 1.6 A and the ion trapping times were recorded as we repeatably trapped and then dumped the ion. We found the time to trap a single $^{138}$Ba$^+$ ion from the BaMg source is on average 49$\pm$7 s.

\section{Discussion}

 In our barium ion trapping tests, when comparing the metallic Ba source with the BaMg sample, there was a discrepancy in the ion loading rates. We attribute this in part to the oven alignment and construction. The ovens were made in-house and while the design was nearly identical, the exact construction may have not been. This means the ovens could heat the samples inside at different rates resulting in varied atomic flux and thus ion loading rates. Similarly, while both these ovens are located on the same vacuum feedthrough of our vacuum chamber, their alignment with respect to the trap apparatus is not the same. Any misalignment from the ovens could cause varied atomic flux through the trapping region that would affect the ion loading rate. 
    
When the BaMg sample is heated, both magnesium and barium vapors are both produced, so there is a possibility of the magnesium becoming ionized through a charge exchange process with the ionized barium. This could result in a magnesium ion becoming trapped instead of barium resulting in lower measured Ba ion loading rate. We note, however, that the ionization energy of magnesium of 7.646~eV is significantly higher than that 5.212~eV of barium. In order for magnesium to be ionized through a charge exchange reaction with a barium ion the interaction would need to be endothermic by approximately 2.434~eV. Assuming the atomic beam source is running at 1000~K, the thermal energy of magnesium in the trap is about 0.1~eV, so the charge exchange reaction of this type is very strongly suppressed. 

Although outside of the scope of what was explored in this paper, we believe that BaMg could be used as a source for laser ablation targets. As discussed earlier, laser ablation techniques typically use thin layers of BaCl$_2$ as targets for barium production and the method for depositing materials onto these targets can be difficult \cite{crystallaser, ablationfix}. BaMg could provide a source that does not oxidize and would not require the same deposition methods as current target materials.

\section{Outlook}

Barium is a desirable qubit for quantum computing applications, but its reactivity can make it technically challenging to work with. The successful use of a barium magnesium alloy in resistive ovens as a source of barium demonstrates a practical alternative to metallic barium for ion trap loading. This alloy may also be a feasible alternative for BaCl$_2$ targets used with laser ablation techniques. Beyond barium, the demonstrated viability of an alloy sample for trapping experiments may be extended to other reactive or difficult to handle atomic species relevant to quantum information processing. Alloy based oven sources represent an experimentally accessible route toward more reliable ion loading and support the continued development of trapped ion quantum technologies.
\vspace{0.55em}
\section{Acknowledgements}

This work is supported by the U.S. National Science Foundation award PHY-2308999 and by the National Science Foundation Graduate Research Fellowship under Grant No. DGE-2140004.

\bibliographystyle{apsrev4-1}
\bibliography{references}

\end{document}